\documentclass[preprint,aps,prc,showpacs,nofootinbib]{revtex4}
\usepackage{epsfig}
\usepackage{graphicx,amsmath}

\newcommand\be{\begin{equation}}
\newcommand\ee{\end{equation}}
\newcommand\ba{\begin{eqnarray}}
\newcommand\ea{\end{eqnarray}}

\newcommand\nn{\nonumber}

\newcommand{\br} [1]{ \left( #1 \right) }
\newcommand{\brs}[1]{ \left[ #1 \right] }

\begin{document}
\title{High order radiative corrections for unpolarized electron proton elastic scattering at low $Q^2$}
\author{E.~A.~Kuraev}
\affiliation{\it JINR-BLTP, 141980 Dubna, Moscow region, Russian Federation}
\author{Yu.~Bystritskiy}
\affiliation{\it JINR-BLTP, 141980 Dubna, Moscow region, Russian Federation}
\author{S.~Bakmaev}
\affiliation{\it JINR-BLTP, 141980 Dubna, Moscow region, Russian Federation}
\author{V.~V.~Bytev}
\affiliation{\it JINR-BLTP, 141980 Dubna, Moscow region, Russian Federation}

\author{E.~Tomasi-Gustafsson}
\affiliation{\it DAPNIA/SPhN, CEA/Saclay, 91191 Gif-sur-Yvette
Cedex, France }

\date{\today}

\begin{abstract}
We investigate the effect of high order radiative corrections in unpolarized electron proton elastic scattering and compare with the calculations at lowest order, which are usually applied to experimental data. We show that higher order terms play a role, starting from values of the momentum transfer squared, $Q^2$, larger than the electron mass. Particular attention is devoted to the $\epsilon$ dependence of radiative corrections.
\end{abstract}

\maketitle
\section{Introduction}

Presently, much attention is devoted to electromagnetic nucleon form factors (FFs) due in particular, to new experimental opportunities to extend their measurement at large momentum transfer and/or to achieve larger precision in the full kinematical region. In particular the possibility to apply the polarization method  \cite{Re68} allowed a measurement of the electric FF up to a value of the momentum transfer squared of 5.8 GeV$^2$  \cite{Jo00}. 

In recent works \cite{ETG07,By07}, it was shown that high order radiative corrections (RC) should be properly taken into account as they may change the size of the observables as well as their dependence on the relevant kinematical variables. 

In this paper we investigate in detail the effect of high order radiative corrections, calculated by the lepton structure function (LSF) method \cite{Ku85}, and  compare them to first order calculations. We focus in particular to the region of low momentum transfer.  The structure function method was developed in Ref. \cite{Ku85} and successfully applied in different processes  (\cite{By07} and refs therein) and allows to calculate high order RC with a precision of 0.1\%.

The main contribution to high order terms, in $ep$ elastic scattering, is due to large logarithm contributions, i.e., such contributions which contain $\log (Q^2)/m^2_e$ ($m_e$ is the electron mass).  One can see that, already at $Q^2\sim 1$ GeV$^2$, such terms become large ($\sim$ 15) partially compensating the factor $\alpha/\pi$ which accompanies the emission of an additional photon. 

The importance of the low $Q^2$ region is related in particular to the high precision parity violating experiments which achieve a $10^{-6}$ precision on the observables (ppm). A precise determination of the  electromagnetic FFs is extremely important in order to decrease the error on the strange nucleon FF.
Measurements at JLab \cite{Gi07} and Mainz \cite{Frank}, based either on the polarization method or on the Rosenbluth separation have been recently done or are in preparation.

In order to extract precise information on the hadron structure, it is necessary to carefully correct the electron block for the emitted photons. This is especially true when the experiment is not fully exclusive, but also in this case, radiative corrections have to be applied within the acceptance and the resolution of the detection. 

We compare the effects of higher order RC to the calculation of Maximon and Tjon \cite{Ma00} (partly rederived in Ref \cite{By07}). The work of Ref. \cite{Ma00} is based on a first order calculation of RC, which improves the classical work of Mo and Tsai \cite{Mo69}, used in most of the analysis codes for elastic and inelastic scattering. It will be quoted below as (MT).

In Section I, the LSF formalism is briefly recalled, and the relevant expressions are given and discussed. In Section II the main results will be presented. The different terms will be compared in first and higher order calculations. In Conclusions we do a brief summary, comment the contribution of inelastic channels, and stress the importance of including high order RC in the codes for the experimental analysis.

\section{Formalism}
The differential cross section for $ep$ elastic scattering in one photon approximation can be expressed as a function of two kinematical variables,
$Q^2$ and $\epsilon$, the four momentum $Q^2$ and the polarization $\epsilon$ of the exchanged virtual photon, in the form:
\be
\frac{d\sigma}{d\Omega}
(Q^2,\epsilon)=\frac{\sigma_M}
{\epsilon \rho(1+\tau)}\sigma_{red}(Q^2,\epsilon),
~\sigma_{red}(Q^2,\epsilon)={\tau }G_M^2(Q^2)+\epsilon G_E^2(Q^2),~
\label{eq:eqphi}
\ee
with
\be
\sigma_M=\frac{Z^2\alpha^2\cos^2(\theta/2)}{4E^2\sin^4(\theta/2)},~
Q^2=-4 E E'\sin^2\displaystyle\frac{\theta}{2},
~\tau=\frac{Q^2}{4M^2},
~\frac{1}{\epsilon }=1+2(1+\tau)\tan^2\displaystyle\frac{\theta}{2}.
\label{eq:eqtau}
\ee
where $\sigma_M$ is the Mott's cross section for electron scattering on point-like particles, and the nucleon structure is described by the form factors, $G_E$ and $G_M$. The kinematical variables are expressed as a function of the incident(final) electron energy $E$ (E'), the electron scattering angle $\theta$. Eqs. (\ref{eq:eqphi},\ref{eq:eqtau}) hold for elastic electron scattering on any hadron, with appropriate values of the mass and the charge of the hadron $M$, $Z$. 

It is known \cite{Ku88} that the process of emission of hard photons by initial and scattered electrons plays a crucial role, which results in the presence of the radiative tail in the distribution on the scattered electron energy. The LSF approach extends the traditional calculation of radiative corrections \cite{Mo69}, taking precisely into account the contributions of higher orders of perturbation theory and the role of initial state photon emission. The cross section can be  expressed in terms of LSF of the initial electron and of the fragmentation function of the scattered electron energy fraction:  
\be
d\sigma^{LSF}(Q^2,\epsilon)=
\int_{z_0}^1 dz {\cal D}(z,\beta)
d\tilde\sigma(Q^2_z,\epsilon_z)\left (1+\frac{\alpha}{\pi}K \right ),~\mbox{~with~}
d\tilde\sigma(Q^2_z,\epsilon_z)= \frac{d\sigma^B(Q^2_z,\epsilon_z)}{|1-\Pi(Q^2_z)|^2},
\label{eq:eqy}
\ee
where $d\tilde\sigma((Q^2_z,\epsilon_z)$, in shifted kinematics, is the Born cross section divided by the correction due to the vacuum polarization.
The $z$-dependent kinematical variables, taking into account the change of the electron four momentum, due to photon emission, 
$Q^2_z$, $\epsilon_z$ are calculated from the corresponding ones (Eq. (
\ref{eq:eqtau})), replacing the initial electron energy $E$ by $zE$, which is the energy fraction carried by the electron after emission of one or more collinear photons.

We used for simplicity the notation $d\sigma$ for the double differential cross section:
$d\sigma^{LSF,B}= (d\sigma^{LSF,B}/{d\Omega})^{LSF,B}$, for Born approximation $(B)$ and radiatively corrected $(LSF)$.
The lower limit of integration, $z_0$, is related to the 'inelasticity' cut, $c$, used to select the elastic data, and corresponds to the maximum energy of the soft photon, which escapes the detection:
\be
z_0=\frac{c}{\rho -c(\rho-1)}, 
\label{eq:eqz}
\ee
where  $\rho$ is the recoil factor $\rho=1+(E/M)(1-\cos\theta)$ and $y=1/\rho $ is the fraction of incident energy carried by the scattered electron. In terms of $\rho$, one can write $Q^2={2E^2(1-\cos\theta)}/{\rho}$.

In Eq. (\ref{eq:eqy}) the main role is played by the non singlet LSF: 
\be
{\cal D}(z,\beta)=\frac{\beta}{2}\brs{\br{1+\frac{3}{8}\beta}(1-z)^{\frac{\beta}{2}-1}-
\frac{1}{2}(1+z)} \br{1+O(\beta)},
\label{eq:eq6}
\ee
\be
    \beta=\frac{2\alpha}{\pi}(L-1),~L=\ln\frac{Q^2}{m_e^2},
\label{eq:eq6a}
\ee
$m_e$ is the electron mass. Particularly important is the quantity $L$, called, large logarithm, which is responsible for the large size of the term related to the LSF correction.

The integration in Eq. (\ref{eq:eqy}) requires a careful treatment, as ${\cal D}(z)$ has a singularity for $z=1$. So the integration of any function $\Phi$ gives (see Appendix A in \cite{By07}):
\ba
{\cal I}&=&\int_{z_0}^1D(x)\Phi(x) dx= \nn\\
&=&\Phi(1)\left [ 1-\displaystyle\frac{\beta}{4}\left (2\ln \displaystyle\frac{1}{1-z_0}-z_0-\displaystyle\frac{z_0^2}{2}\right )\right ]
+\displaystyle\frac{\beta}{4}\int_{z_0}^1 dx\displaystyle\frac{1+x^2}{1-x}\left [\Phi(x)-\Phi(1)\right ]+{\cal O}(\beta^2).
\label{eq:eqa1}
\ea
The factor $1+({\alpha}/{\pi })K$ can be considered as a general normalization. It has been calculated in detail for $ep$ elastic scattering in Ref. \cite{By07} and the term $K$ is the sum of three contributions:
\be
K=K_e+K_p+K_{box}.
\label{eq:eqkfac}
\ee
$K_e$ is related to non leading contributions arising from the pure electron block  and can be written as \cite{Ku85,Ku88}:
\be K_e= -\displaystyle\frac{\pi^2}{6} -\displaystyle\frac{1}{2} -\displaystyle\frac{1}{2}\ln^2\rho+Li_2(\cos^2\theta/2),~
Li_2(z)=-\int_0^z \displaystyle\frac{dx}{x} \ln(1-x).
\label{eq:eqll}
\ee
The second term, $K_p$,  concerns the emission from the proton block. The emission of virtual and soft photons by the proton is not associated with large logarithm, $L$, therefore the whole proton contribution can be included as a $K_p$ factor:
\begin{eqnarray}
K_p&=&\displaystyle\frac{Z^2}{\beta}\left \{ -\displaystyle\frac{1}{2}\ln^2x-\ln x\ln [4(1+\tau)]
+\ln x - \right .
\nonumber \\
&&\left . (\ln x-\beta)\ln\left [ \displaystyle\frac{M^2}{4E^2(1-c)^2}\right ]+\beta -Li_2\left (1-\displaystyle\frac{1}{x^2}\right )+2Li_2\left(-\displaystyle\frac{1}{x}\right )+
\displaystyle\frac{\pi^2}{6} \right\},
\label{eq:eqkp}
\end{eqnarray}
with $x=(\sqrt{1+\tau}+\sqrt{\tau})^2$, $\beta=\sqrt{1-M^2/E'^2}$ and $E'=E(1-1/\rho)+M$ are the scattered proton velocity and energy.
The contribution of $K_p$ to the $K$ factor is of the order of -.2\% for $c=0.99$, $E=21.5$ GeV, $Q^2$=31.3 GeV$^2$ \cite{Ma00}, and it is almost constant in $\epsilon$.  

Lastly, $K_{box}$ represents the interference of electron and proton emission. More precisely the interference between the two virtual photon exchange amplitude and the Born amplitude as well as the relevant part of the soft photon emission i.e., the interference between the electron and proton soft photon emission, may be both included in the term $K_{box}$. These effects are not enhanced by large logarithm (characteristic of LSF) and can be considered among the non-leading contribution, which represents an $\epsilon$-independent quantity of the order of unity, including all the non-leading terms, as two photon exchange and soft photon emission.
 
In order to make comparison with existing calculations of RC, it is convenient to express the corrections calculated with the LSF method, $\delta $ in the form:
\be 
d\sigma^{LSF}(Q^2,\epsilon)=d\sigma^B(Q^2,\epsilon)(1+\delta),
\label{eq:csa}
\ee 
where 
\ba 
1+\delta&=&
\frac{1}{|1-\Pi(Q^2)|^2}
\left \{ 
1+\displaystyle\frac{\alpha}{2\pi}(L-1)
\left [
-\left (2\ln\left (\displaystyle\frac{1}{1-z_0}\right )-z_0-\displaystyle\frac{z_0^2}{2}
\right )+
\right .\right .
\nn\\
&&
\left . 
\int_{z_0}^1 dz\displaystyle\frac{1+z^2}{1-z}
\left .
\left(
\displaystyle\frac
{d\tilde\sigma^B(Q^2_z,\epsilon_z)}
{ d\tilde\sigma^B(Q^2,\epsilon)}-1\right ) \right ]
+\displaystyle\frac{\alpha}{\pi} K.\right \}
\label{eq:cs}
\ea
Let us compare these different terms with the corresponding calculation from Ref. \cite{Ma00}, which has removed or softened some drastic approximations previously used in \cite{Mo69}. The interference between the box and the Born diagram was included (partially within the soft photon approximation) as:
\be
\delta^{box}=\frac{2\alpha }{\pi}Z\left \{ -\ln\rho\ln
\left [\frac{-q^2x}{(2\rho\Delta E)^2} \right ]
+ Li_2\left (1-\frac{\rho }{ x}\right )
- Li_2\left (1-\frac{1}{\rho x}\right )\right \},
\label{eq:MTbox}
\ee
where $\Delta E= E'(1-c)$ is the maximum energy of the soft photon, allowed by the experimental set-up. 
The radiation from the electron, in the leading order approximation, including vacuum polarization, was expressed as 
\be
\delta^{el}=\frac{\alpha }{\pi}\left \{ \frac{13}{6}L 
 - \frac{28}{9} -( L -1 ) \ln\left [\frac {4 EE'}{(2\rho \Delta E)^2}\right ] 
-\frac{1}{2}\ln^2\rho +Li_2\left (\cos^2\frac{\theta}{2}\right ) 
-\frac{\pi^2}{6}\right \}. 
\label{eq:MTel}
\ee

\section{Results and Conclusions}
In the LSF calculation, the main contribution comes from those terms, which include higher order corrections, whereas all the terms which do not contain large logarithm are expected to be suppressed and included in the $K$-factor.
 
\begin{figure}
\begin{center}
\includegraphics[width=12cm]{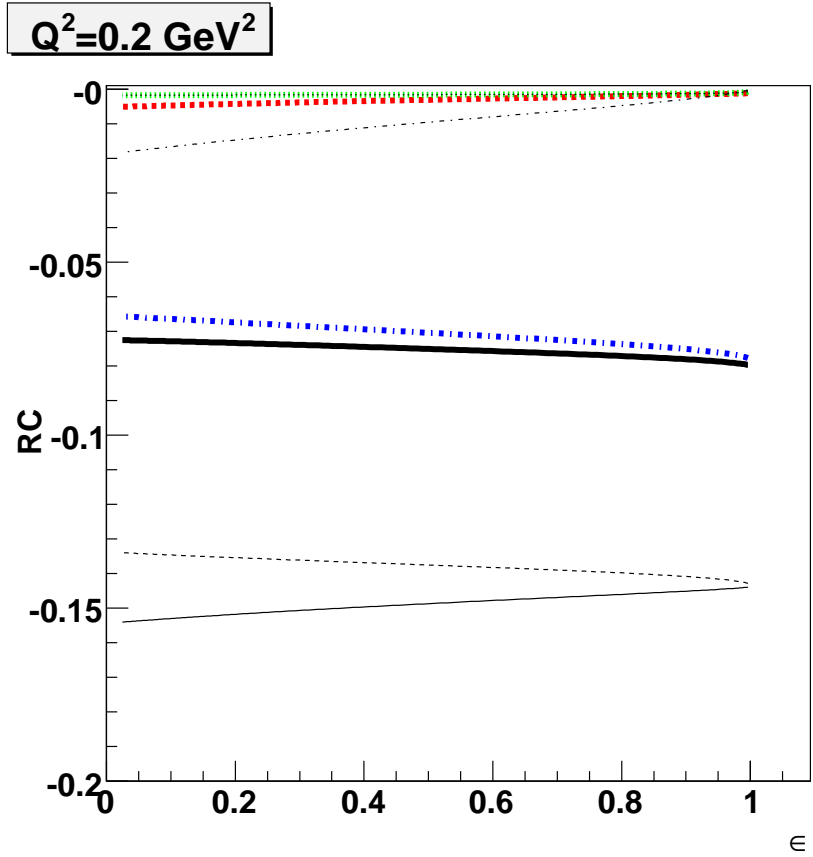}
\caption{\label{Fig:fig1}(Color online)Calculation of different radiative corrections for $Q^2$=0.2 GeV$^2$. 
Thick lines correspond to LSF calculation, thin black lines correspond to MT \protect\cite{Ma00}. Total correction (solid lines), electron emission (red and black dashed lines), proton emission 'green and black dotted lines), Structure function (LL) (blue dash-dotted line), box MT (black, thin, dash-dotted line)  }
\end{center}
\end{figure}

In Fig. \ref{Fig:fig1}  the results for the calculation of different radiative corrections for $Q^2$=0.2 GeV$^2$ are shown. Thick lines correspond to the LSF method and thin lines correspond to MT \cite{Ma00}. The solid line is the sum of the different terms. A large difference can be already seen at such small value of $Q^2$, both in the values and in the slope of the cross section. 

Let us compare the different terms. 

The correction from the proton (dash-dotted line) is basically the same, in both calculations (Eq. \ref{eq:eqkp}). It is small and $\epsilon$ independent. 

For both methods, the largest contribution is due to the radiation from the electron. However, in the LSF method, the main correction is due to the electron radiation ( dash-dotted line), whereas the corrections from the electron which do not contain large logarithm and are  calculated in the $K_e$ factor, which is small (dashed line), with a small $\epsilon$ dependence. 

The electron emission from the MT calculation, Eq. (\ref{eq:MTel}) is shown as a thin, dashed line, and corresponds to the largest contribution to RC.
In the LSF method, the emission from the initial electron is taken into account. For the  final electron emission, it has been assumed that the full energy is detected (for example, if the electron is detected in a calorimeter) or that the electron energy is not measured at all: in these cases, due to the properties of LSF, the contribution for final emission is unity.

For the LSF calculation, here we take $K_{box}=0$.  In Refs. \cite{By07,EAK08} it was shown that this term was small, and with small $\epsilon$ dependence. 

In the MT calculation, the interference between the box and the born diagram (thin,black dash-dotted line) has a positive slope, and a large  $\epsilon$ dependence (Eq. \ref{eq:MTbox}). Therefore it is this term which is responsible for the fact that the slopes of the final corrections as a function of $\epsilon$ (thin and thick solid lines ) have opposite signs in the LSF and MT calculations. When applied to the experimental cross sections, this will be reflected in a change of slope of the reduced cross section, as a function of $\epsilon$, and the electromagnetic FFs extracted from the Rosenbluth method will be different. In Ref. \cite{By07} it has been shown that the LSF corrections could bring into agreement the FFs extracted by the Rosenbluth and the polarization methods.  
 
\begin{figure}
\begin{center}
\includegraphics[width=12cm]{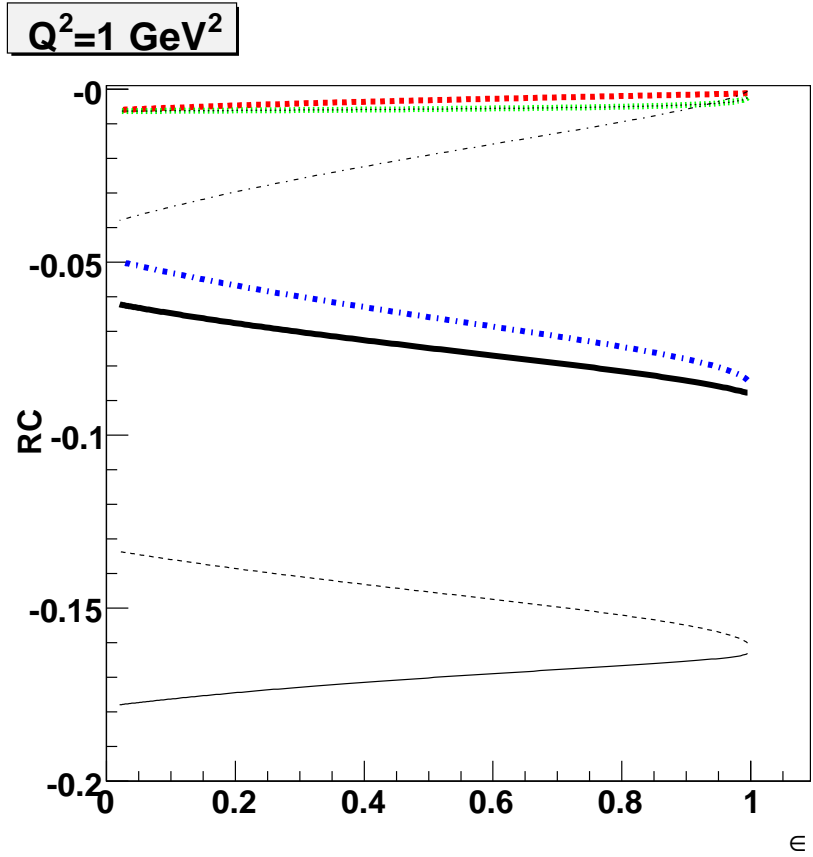}
\caption{\label{Fig:fig2}(Color online)Same as Fig. \protect\ref{Fig:fig1}, for $Q^2$=1 GeV$^2$.  }
\end{center}
\end{figure}

In Fig. \ref{Fig:fig2} the different contributions to RC are shown for $Q^2$=1 GeV$^2$, as a function of $\epsilon$. The general behavior of the different terms is essentially similar as in Fig.  \ref{Fig:fig1}, but the numerical values of the radiative corrections are larger and the effect of higher orders more sizable. Numerical values may differ up to a factor of three, and this factor depends on $\epsilon$ and $Q^2$. 
 
\begin{figure}
\begin{center}
\includegraphics[width=12cm]{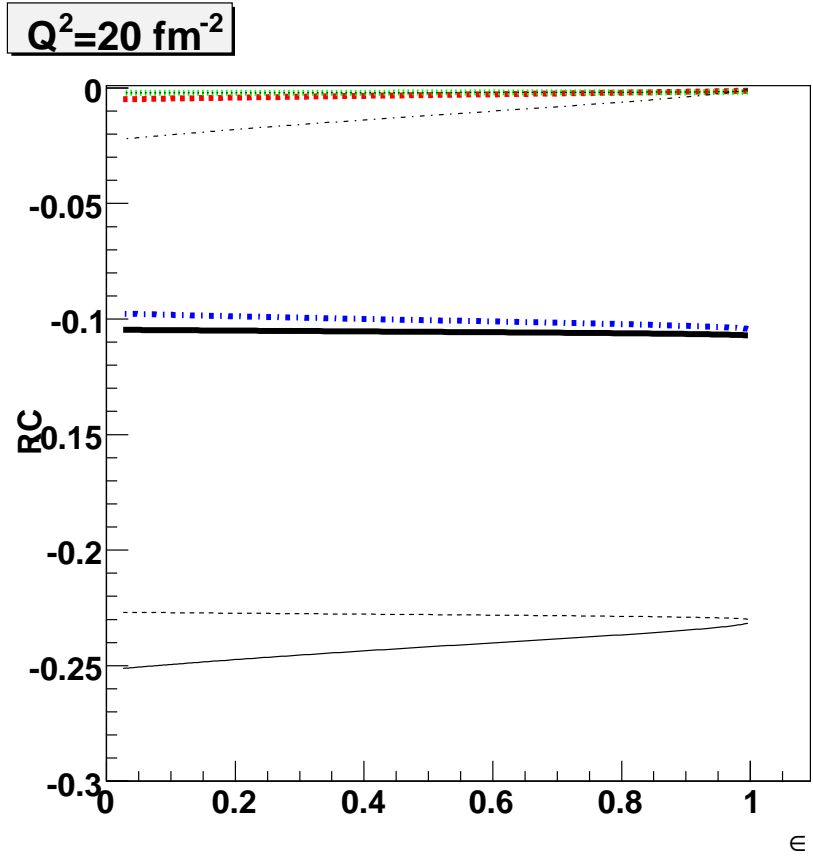}
\caption{(Color online)Calculation of different radiative corrections for $ e+^4\!He$ elastic scattering, at $Q^2$=0.8 GeV$^2$. (notations as in Fig. \protect\ref{Fig:fig1}}
\label{Fig:rcHe} 
\end{center}
\end{figure}

In Fig. \ref{Fig:rcHe} the results are shown in case of Helium target, at $Q^2$=0.8 GeV$^2$. 

To summarize, the main difference between the size of RC from the two calculations should be attributed to the fact that in the present application of the LSF method, the partition function of the final particle is taken as unity, which is the case in an experiment where one can not separate events corresponding to an electron and to an electron and a photon with the same total energy. The difference between the slopes of the corrections to the cross section as a function of $\epsilon$  depends on the ansatz used to include 
the two photon exchange mechanism.
\section{Conclusions}

We have calculated radiative corrections for electron hadron elastic scattering, at low $Q^2$, in frame of the LSF method, and compared to lowest order calculations. Even if a comparison can not be done term by term, as the formalisms are different, we can draw the following conclusions.

Radiative corrections by the LSF method are in general of the same sign, negative, but smaller than for MT, and they have the effect to increase the cross section, when compared to the calculations at the lowest order. The two calculations should basically agree at the lowest order of PT. The difference between the two calculations comes mainly from the fact that, in this application of LSF approach, the contribution of the final emission is unity and that higher order are taken into account (in the leading logarithm approximation).

Let us discuss one of possible source of non-leading contributions to
elastic-proton scattering -- the interference of one and two photon
exchange amplitudes. The structure of the proton can be taken into account
by introducing a generalized form factor, which describes the interaction of
an off-mass-shell photon with an on-mass-shell nucleon with production of
an off-mass-shell proton or of some inelastic state (set of nucleons, antinucleons, and pions). Arguments of analyticity
and gauge invariance can be applied to the Compton amplitude
$\gamma^* p \to \gamma^* p$. The integration over the  invariant mass of the initial particles leads to the formulation of "sum rules" and to the statement 
that the contributions of one-proton intermediate state in Compton amplitude and the inelastic ones essentially compensate each other \cite{Kuraev:2007pv,Kuraev:2006ym}.
The relevant contribution of the interference of one and two photon
exchange amplitudes can be included as a $K$-factor, and, in the case of small $Q^2$, it is of the order:
\be
     K_{\gamma\gamma} = 1 +
     \frac{\alpha}{\pi} \frac{Q^2}{(3M_p)^2}.
\ee
For these reasons, we neglect this quantity in the present estimation of RC. Moreover, explicit calculations of elastic and inelastic contributions
\cite{Borisyuk:2006fh,Blunden:2005ew}, in frame of models, also support the statement of the discussed compensation.

The different sign of the slope of RC as a function of $\epsilon$ comes, in the MT calculation, from corrections due to the box diagram. We remind that, in this calculation, one photon is soft and the other hard, and the corresponding terms are introduced in order to compensate the infrared divergence due to soft photon emission. Indeed, it was shown in \cite{Ku06} in an exact QED calculation for $e\mu$ scattering, that the box contribution is very small (few thousandth). On the other hand, the charge asymmetry, in the reaction $e^+ +e^-\to \mu^+ +\mu^-$ can be measurable, of the order of percent, due to the contribution of soft photon emission. Similar conclusions hold in case of Helium target.


\end{document}